\documentclass[twocolumn]{aastex63}
\usepackage{graphicx}
\usepackage{subfigure}
\usepackage{color, hyperref, epsfig}
\usepackage{apjfonts, natbib}
\usepackage{appendix}
\usepackage{float}
\usepackage{bm}

\newcommand{\lum}{erg~s\ensuremath{^{-1}}}

\newcommand{\mbh}{\ensuremath{M_\mathrm{BH}}}

\shorttitle{Obscured TDEs with Neutrinos}
\shortauthors{Jiang et al.}

\begin{document}

\title{Two Candidate Obscured Tidal Disruption Events Coincident with High-energy Neutrinos}
\correspondingauthor{Ning Jiang}
\email{jnac@ustc.edu.cn}

\author[0000-0002-7152-3621]{Ning Jiang}
\affiliation{CAS Key laboratory for Research in Galaxies and Cosmology,
Department of Astronomy, University of Science and Technology of China, 
Hefei, 230026, China; jnac@ustc.edu.cn}
\affiliation{School of Astronomy and Space Sciences,
University of Science and Technology of China, Hefei, 230026, China}

\author[0009-0006-5973-4043]{Ziying~Zhou}
\affiliation{CAS Key laboratory for Research in Galaxies and Cosmology,
Department of Astronomy, University of Science and Technology of China, 
Hefei, 230026, China; 
jnac@ustc.edu.cn}
\affiliation{School of Astronomy and Space Sciences,
University of Science and Technology of China, Hefei, 230026, China}

\author[0000-0003-3824-9496]{Jiazheng~Zhu}
\affiliation{CAS Key laboratory for Research in Galaxies and Cosmology,
Department of Astronomy, University of Science and Technology of China, 
Hefei, 230026, China; 
jnac@ustc.edu.cn}
\affiliation{School of Astronomy and Space Sciences,
University of Science and Technology of China, Hefei, 230026, China}

\author[0000-0003-4225-5442]{Yibo~Wang}
\affiliation{CAS Key laboratory for Research in Galaxies and Cosmology,
Department of Astronomy, University of Science and Technology of China, 
Hefei, 230026, China; 
jnac@ustc.edu.cn}
\affiliation{School of Astronomy and Space Sciences,
University of Science and Technology of China, Hefei, 230026, China}

\author[0000-0002-1517-6792]{Tinggui~Wang}
\affiliation{CAS Key laboratory for Research in Galaxies and Cosmology,
Department of Astronomy, University of Science and Technology of China, 
Hefei, 230026, China; 
jnac@ustc.edu.cn}
\affiliation{School of Astronomy and Space Sciences,
University of Science and Technology of China, Hefei, 230026, China}

\begin{abstract}
Recently, three optical tidal disruption event (TDE) candidates discovered by the Zwicky Transient Facility (ZTF) have been suggested to be coincident with high-energy neutrinos. They all exhibit unusually strong dust infrared (IR) echoes, with their peak times matching the neutrino arrival time even better than the optical peaks. We hereby report on two new TDE candidates that are spatially and temporally coincident with neutrinos by matching our sample of mid-infrared outbursts in nearby galaxies (MIRONG) with Gold alerts of IceCube high-energy neutrino events up to June 2022. The two candidates show negligible optical variability according to their ZTF light curves and can therefore be classified as part of the growing population of obscured TDE candidates. The chance probability of finding two such candidates about $\sim3\%$ by redistributing the MIRONG sources randomly in the SDSS footprint,  which will be as low as $\sim0.1\%$ (or $\sim0.2\%$) if we limit to sources with increased fluxes (or variability amplitudes) comparable with the matched two sources. Our findings further support the potential connection between high-energy neutrinos and TDEs in dusty environments by increasing the total number of neutrino-associated TDE and TDE candidates to five, although the underlying physics remains poorly understood.

\end{abstract}

\keywords{Tidal disruption (1696) --- Neutrino astronomy (1100) --- Supermassive black holes (1663) --- High energy astrophysics (739) --- Time domain astronomy (2109)}

\section{introduction}

Since the detection of TeV-PeV cosmic neutrinos by the IceCube neutrino observatory in 2013 \citep{IceCube2013}, the number of high-energy neutrinos has steadily increased. However, their astrophysical origin remains a hotly debated topic and is one of the most cutting-edge areas of research in multi-messenger astronomy. Despite the increasing number of high-energy neutrinos detected, no significant cluster in space or time has been found while there is a tentative correlation with the nearby Seyfert galaxy NGC~1068 at the 4.2$\sigma$ level~\citep{Aartsen2020,IceCube2022}. Therefore, identifying electromagnetic counterparts of high-energy neutrinos remains extremely challenging.

A complementary approach is to directly search for electromagnetic counterparts to individual high-energy neutrinos that have a high probability of being of astrophysical origin. However, only three candidates of electromagnetic counterparts have been identified at the $\sim3\sigma$ level to date based on both spatial and temporal coincidence. The first ever identification was the flaring blazar TXS~0506+056~\citep{IceCube2018} with neutrino alert IC170922A during a period of electromagnetic flaring in 2017. The latter two were identified as a tidal disruption event (TDE, \citealt{Gezari2021}) and a TDE candidate, respectively. TDEs occur when an unlucky star wanders into the tidal sphere of a SMBH and gets ripped apart by the tidal force. Part of the disrupted stellar debris is accreted by the SMBH and produces an electromagnetic flare on timescales of months to years~\citep{Rees1988}. The discovery speed of TDEs has accelerated with the aid of modern optical time-domain surveys, particularly the Zwicky Transient Facility (ZTF), which has boosted the discovery rate from $\lesssim2/\rm yr$ per year to $>10/\rm yr$~\citep{vV2021apj,Yao2023}. Both objects, AT2019dsg and AT2019fdr, were initially discovered by ZTF, and their optical luminosity peaks occurred 150-300 days before the arrival time of the corresponding IceCube neutrino events, that is IC191001A (\citealt{Stein2021}, cf. \citealt{Liao2022}) and IC200530A~(\citealt{Reusch2022}, cf. \citealt{Pitik2022}), respectively. AT2019dsg is a bonafide TDE and its property has been detailedly analyzed by works on ZTF TDE sample~\citep{vV2021apj,Hammerstein2023,Yao2023}. In contrast, the TDE identification of AT2019fdr is rather uncertain as the flare happened in a known AGN, the origin of which is complicated by the existence of accretion disk (e.g., \citealt{Graham2017,Trakhtenbrot2019,Frederick2021}). The long-lived non-thermal emission detected in AT2019dsg~\citep{Stein2021,Cendes2021} indicates that the mildly relativistic outflows may be an ideal site for neutrino production. Both of the two neutrino emitters show unusually strong infrared (IR) echoes, which is the reprocessed emission of high-energy photons by the dust in the vicinity of the SMBHs~\citep{Jiang2016,vV2016}. Moreover, it is interesting to note that the arrival time of the neutrinos matches the peak time of the IR emission even better than the peak time of the optical emission.

Actually, most optical TDEs show very weak IR echoes, suggesting a dust-poor subparsec environment~\citep{Jiang2021apj}. The unusual properties of the neutrino-associated TDEs (or candidates) motivated a systematic search for similar nuclear transients with prominent IR echoes, and a third similar object AT2019aalc, occuring in another known AGN, was found as the counterpart of IC191119A~\citep{vV2021b}. Moreover, they obtained a correlation at the $3.6\sigma$ confidence level between such flares and high-energy neutrino alerts. However, it should be noted that their search primarily started with the selection of optical flares in galactic nuclei with ZTF light curves, and as a result, all of these events are optically bright and presumably dust-unobscured. Nevertheless, an independent systematic search for mid-IR outbursts in nearby galaxies (MIRONG) suggests that a large fraction of nuclear transients might be optically weak~\citep{Jiang2021apjs}. Therefore, if high-energy neutrinos are indeed connected with nuclear transients with strong IR echoes, as claimed by~\citet{vV2021b}, the most efficient search for their counterparts should rely on the IR-selected sample, which is dominated by transient accretions onto SMBHs in dusty environments~\citep{Jiang2021apjs, Wang2022apjs}, including unambiguous TDEs~\citep{Wang2022apjl}. Most importantly, such TDEs can only be identified by IR echoes or radio emission if they are severely dust-obscured (e.g., \citealt{Mattila2018,Kool2020,Panagiotou2023}).

In this letter, we report the discovery of two obscured TDE candidates in the MIRONG sample that coincide spatially and temporally with high-energy neutrinos. We assume a cosmology with $H_{0} =70$ km~s$^{-1}$~Mpc$^{-1}$, $\Omega_{m} = 0.3$, and $\Omega_{\Lambda} = 0.7$.

\section{Data}

\begin{figure*}
\centering
\begin{minipage}{1.0\textwidth}
\centering{\includegraphics[width=1\textwidth]{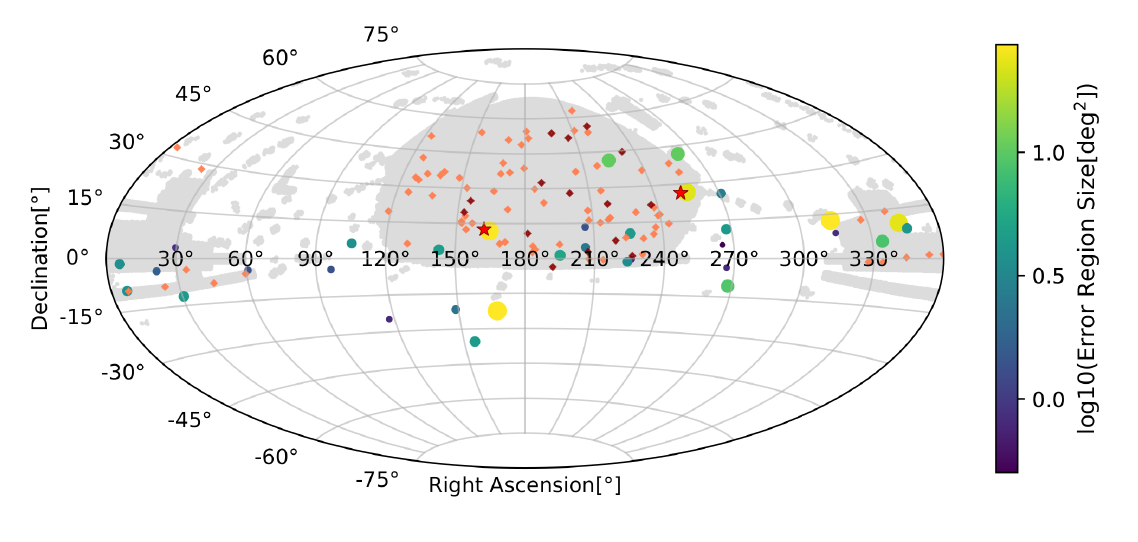}}
\end{minipage}
\caption{Sky map of observed high-energy neutrino event (Gold alerts) positions and their corresponding error region sizes, which are indicated by a colorbar on the right. The footprint of SDSS, from which our MIRONG sample is selected, is overplotted in grey overlay.
The objects in our MIRONG sample with IR peaks at the same period are denoted as red diamonds and those with $W2\leq12.60$ are marked as dark red. The two sources that coincide with neutrinos are highlighted as red five-point stars. The clustering effect of the events around the equator region is due to the detector's high sensitivity in this area.\label{map}
}
\end{figure*}

\subsection{The Neutrino Events}
The IceCube Neutrino Observatory has detected high-energy astrophysical neutrinos that are statistically significant above the atmospheric background. Since 2016, real-time alerts have been issued to the multi-messenger observational community for single high-energy ($>$100~TeV) events via platforms such as the Global Cycling Network (GCN)~\footnote{https://gcn.gsfc.nasa.gov/amon\_icecube\_gold\_bronze\_events.html}. These alerts focus on track-like neutrino candidates, which offer more accurate angular localizations than cascade events. All alerts since June 2019 are assigned a signalness value and marked as gold or bronze depending on the chance of astrophysical origin, i.e., larger than 50\% and 30\%, respectively. The 90\% containment error radius ("ERROR90") of these events ranges from 0.4 to 6.2 degrees, with an average of 1.5 degrees, not including systematic error. Note that for most neutrinos, a more accurate position is provided after the original one. These updated positions and their corresponding 90\% containment errors are adopted in our analysis~\footnote{We also note that there is a recent paper~\citep{Abbasi2023} which has provided updated 90\% containment for neutrinos between 2011 and 2020. However, we are mostly interested in the neutrinos between 2019 and 2022 since they are well overlapped with the ZTF survey timeline, which can offer crucial information on their optical emission and thus help diagnose their nature.}.
The occurrence rate is expected to be one per month. 
For our analysis, we use the list of gold alerts between June 2019 and June 2022. Moreover, we only consider neutrinos with "ERROR90" less than 200 arcmin, which finally includes a total of 33 events (see Figure~\ref{map} and a full list in Table~\ref{tb-neut} in the Appendix).

\subsection{The parent MIRONG Sample}
The objects we match with high-energy neutrinos are from the MIRONG sample, which was constructed by a systematic search of low-redshift ($z<0.35$) Sloan Digital Sky Survey (SDSS) spectroscopic galaxies with recent mid-IR flares. These objects were identified using archival data from the Wide-field Infrared Survey Explorer (WISE, \citealt{Wright2010}) and its new mission, the Near-Earth Object WISE (NEOWISE, \citealt{Mainzer2014}). The original sample, using data up to the end of 2018, resulted in a total of 137 objects with a brightening amplitude of 0.5 mag in at least one band with respect to their quiescent phases~\citep{Jiang2021apjs}. In the current work, we extend the same search to the data as of the end of 2022, which is the most recent public NEOWISE dataset. The new MIRONG sample contains 269 objects, and its general properties remain the same as reported in \citet{Jiang2021apjs}.

\begin{figure*}
\centering
\begin{minipage}{1.0\textwidth}
\centering{\includegraphics[width=1\textwidth]{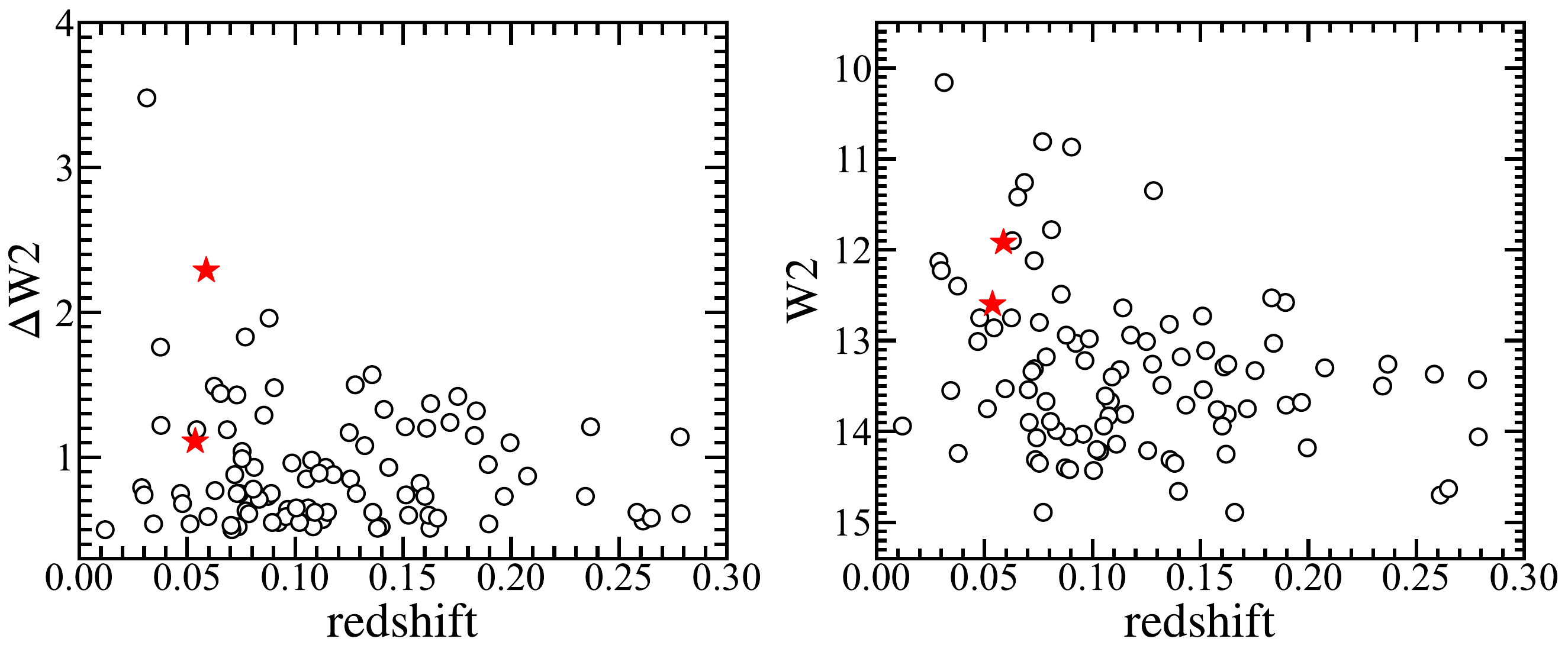}}
\end{minipage}
\caption{Left: the W2 variability amplitude ($\Delta W2$) vs. redshift distribution of the MIRONG sources used to match with neutrinos. Right: the W2 magnitude vs. redshift distribution. The two sources coincident with neutrinos are highlighted by red five-point stars.\label{dis}
}
\end{figure*}

\section{Analysis and Results}

\subsection{Cross-match}
\label{crossmatch}
The IR light curves for each object have been continuously sampled every six months in the WISE W1 ($3.4\mu$m) and W2 ($4.6\mu$m) bands since 2014. In order to temporally crossmatch potential sources with neutrinos, we only select those that exhibit peaks between June 2019 and June 2022, coinciding with the neutrino time range. The sources that are still in the rising stages as of the latest epoch (after June 2022) have been excluded. This process results in a total of 94 sources, which is listed in Table~\ref{tb-mirong} in the Appendix. For each IceCube high-energy neutrino, there are its arrival time, NoticeType ("GOLD" or "BRONZE"), location (R.A., Decl.), energy, and signalness. Two location uncertainties, one with 90\% ("ERROR90") and the other with 50\% ("ERROR50") containment, are assigned. We only considered gold events and used the "ERROR90" location uncertainty as the matching radius. Furthermore, we define a temporal match by requiring that the time interval between the IR brightest epoch and the neutrino arrival is less than half a year, which is the WISE visit cadence.

Following above procedures, two outbursts in the MIRONG sample show both spatial and temporal coincidences with gold neutrinos. They are SDSSJ104832.79+122857.2 (hereafter SDSSJ1048+1228) and SDSSJ164938.77+262515.3 (SDSSJ1649+2625), matching with IceCube-200109A and IceCube-200530A, respectively (see their information in Table~\ref{twoobj}). Specifically, the MIR outburst in SDSSJ1048+1228 peaks on 2020 May 8 (MJD=58977) in WISE W2 band, which is four months ahead of IceCube-200109A while IceCube-200530A arrives about 2 months earlier than the IR peak time of SDSSJ1649+2625 on 2020 August 7 (MJD=59068). Spatially, the two neutrinos are 159 and 143 arcmins away from the two galaxies, yet they are consistent with each other within the 90\% location uncertainties.

\begin{deluxetable}{c|cc}
\tablecaption{Information of the matched MIR outbursts and neutrinos.\label{twoobj}}
\tablehead{ & SDSSJ1048+1228 & SDSSJ1649+2625 }
\startdata
redshift     & 0.0537      & 0.0588    \\
BPT Type   &  LINER   &   Composite \\
Host Stellar mass (log$M_{\star}$) & 10.86 & 9.13  \\
log$\mbh^{a}$ &  7.86  &  6.49 \\
IR peak time & 2020/05/08  &  2020/08/07 \\
$\Delta W1/\Delta W2$ & 0.59/1.11  &  1.41/2.29   \\
$\rm W1/W2~(peak magnitude)^{b}$ &  13.65/12.60  & 13.09/11.92  \\
IR peak luminosity & 42.84  &  43.20 \\
IR energy   & 50.72    &  51.02  \\
\hline
Neutrino     & IceCube-200109A &  IceCube-200530A  \\
NoticeType   & GOLD & GOLD \\
Arrival Date & 2020/01/09  & 2020/05/30  \\
ERROR90      & 174.60      & 160.19  \\
Energy       & 375.23      & 82.186  \\
\enddata    
\tablecomments{\\
$\rm ^{a}$ The mass of SMBH is estimated from $M_{\rm BH}-\sigma_{\star}$ relation following that in~\citet{Jiang2021apjs}. \\
$\rm ^{b}$ The magnitude is for the emission with the quiescent-state flux subtracted.
}
\end{deluxetable}

\begin{figure}
\centering
\begin{minipage}{0.5\textwidth}
\centering{\includegraphics[width=1\textwidth]{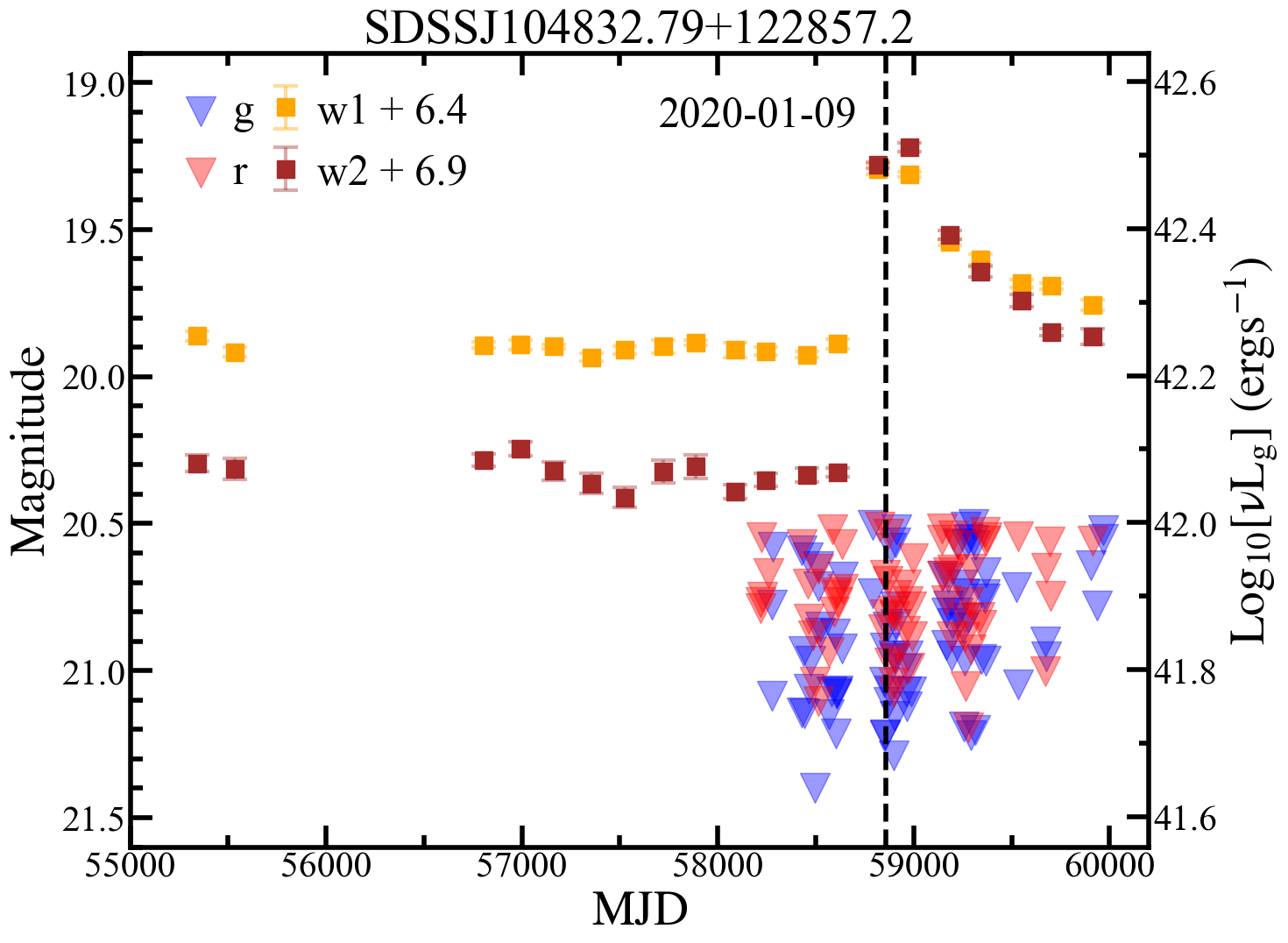}}
\centering{\includegraphics[width=1\textwidth]{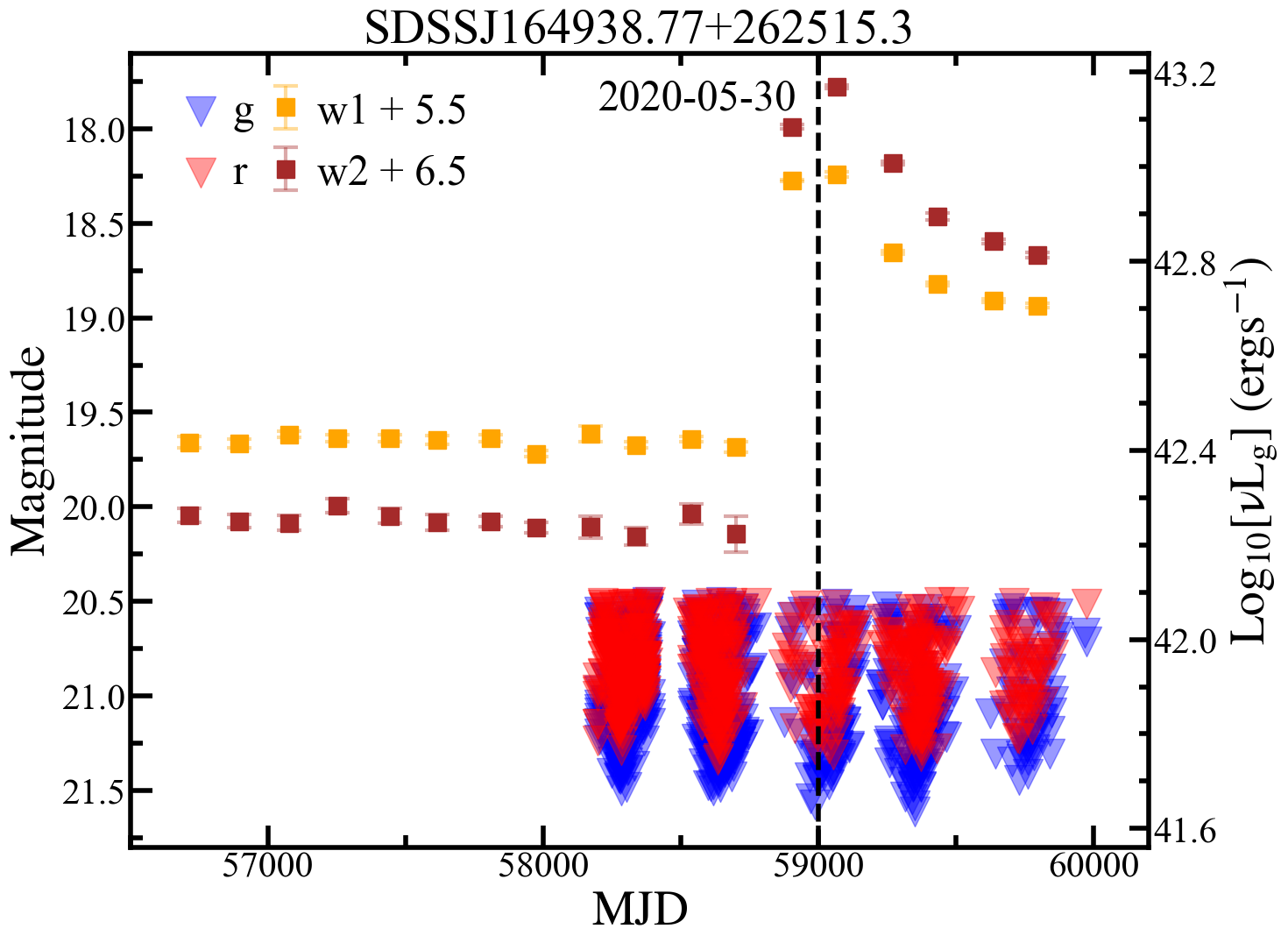}}
\end{minipage}
\caption{The MIR and optical light curves of SDSSJ1048+1228 and SDSSJ1649+2625. The NEOWISE photometry in W1 ($3.4\mu$m) and W2 ($4.6\mu$m) band are shown as orange and brown squares while the blue and red triangles represent the $3\sigma$ upper limits of ZTF forced difference photometry in $g$ and $r$ bands.
The vertical dashed lines mark the arrival time of matched high-energy neutrinos IceCube-200109A and IceCube-200530A.\label{lc}
}
\end{figure}

\subsection{Probability by chance}
We then attempt to calculate the probability of finding two such coincident events by chance. 
First, we perform a simple estimation following the method used in \citet{Stein2021}. The area of the SDSS footprint and the total 90\% containment area of the neutrinos in the SDSS footprint are $9376~\rm deg^2$ and $127.8~\rm deg^2$, respectively. There are 94 MIRONG sources peaked between June 2019 and June 2022, thus their number density is $94/9376/3=3.3\times10^{-3}/\rm deg^2/yr$. Since we match the MIRONG sources with neutrinos by requesting their intervals are less than half year, the matching time window is thus 1 yr. The expectation value for the number of random matches is thus $3.3\times10^{-3}\times127.8=0.427$, and then the probability of observing two or more matches is 0.069 based on Poisson statistics.

It is worthwhile to note that SDSSJ1048+1228 and SDSSJ1649+2625 belong to the lower-redshift (top 12\%), larger amplitude (top 28\%), and more apparently luminous (top 18\%) subclass among the all sources (see Figure~\ref{dis}), indicating that the neutrino emission is likely related to a special population of MIRONG objects. One simple assumption is that the neutrino count is proportional to the observed MIR flux and thus the neutrino counterparts favour brighter MIR sources. If we restrict the sources to those with an increased flux comparable with the matched two, i.e. $W2<12.60$, the number of sources decreases from 94 do 17. As a result, the corresponding chance probability of finding two matches is calculated to be 0.0028. The other scenario is that the neutrinos are correlated with the IR variability amplitudes, as explored in the analysis conducted by \citet{vV2021b}. By limiting the MIRONG sources to those with variability amplitudes similar to the matched two, i.e. $\Delta W2>1$, the number of sources becomes 30 and the chance probability is then 0.0085.

We also try to estimate the probability by simulations.
It should be noted that the flux limits of IceCube show a strong dependence on decl.~\citep{Aartsen2020}. To account for this effect, we perform Monte Carlo simulations by redistributing the 33 gold neutrinos across sky taking into account the declinational dependence  while with their distribution in right ascension entirely random. Specifically, the declination was redistributed according to the actual probability distribution
\footnote{see Figure~3 in  \url{https://gcn.gsfc.nasa.gov/doc/IceCube\_High\_Energy\_Neutrino\_Track\_Alerts\_v2.pdf}. We did not employ a parametric function to model the distribution. Instead, we simply assumed that the possibility within each histogram is evenly distributed.}. Furthermore, we did not alter the arrival time of neutrinos in our simulations since the timing of neutrino arrivals is random and not expected to have a significant impact on our results. 
We then match them spatially and temporarily with the MIRONG sources as described in \S\ref{crossmatch}. The experiment is repeated 30,000 times. The resulting probability of finding at least two matches is 0.0296 for 94 sources, which decreases to 0.0013 for 17 sources ($W2<12.6$) and decreas from 0.0049 for 30 sources ($\Delta W2>1$). 

Another alternative simulation involves shuffling the MIRONG sources while keeping the neutrinos fixed. This approach could be more reliable due to a relatively constant source density within the SDSS footprint, while the sky and temporal sensitivities of IceCube are challenging to quantify accurately. Therefore, we randomly redistributed the MIRONG sources within the SDSS footprint and subsequently matched them with neutrinos. The resulting chance probabilities of finding at least two matches are 0.025 and 0.0009 in the case without and with the requirement of $W2<12.6$, respectively, which are slightly lower than the case of shuffling neutrinos. Additionally, the probability is reduced to 0.0023 for the 30 sources with $\Delta W2>1$. 
 
\subsection{The Two Matched Sources as Obscured TDE candidates}
The two MIR outbursts potentially associated with high-energy neutrinos lack detectable optical counterparts in public surveys such as ZTF (\citealt{Masci2019}, see Figure~\ref{lc}). In fact, as noted by \citet{Jiang2021apjs}, the majority of the outbursts in the MIRONG sample have not been previously reported by any other surveys. This indicates that their optical emissions are either obscured or intrinsically weak.

\begin{figure}
\centering
\begin{minipage}{0.5\textwidth}
\centering{\includegraphics[width=1\textwidth]{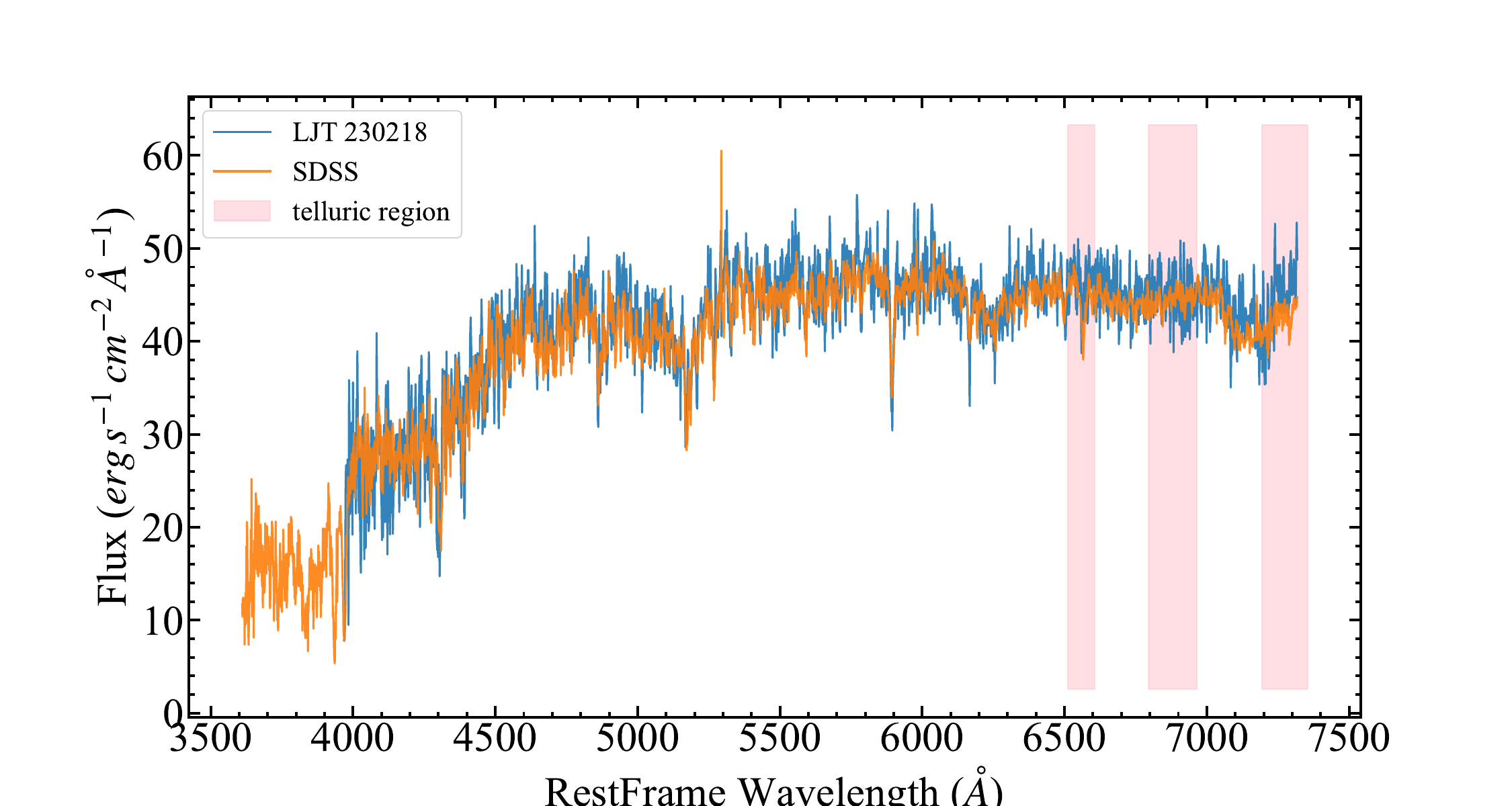}}
\centering{\includegraphics[width=1\textwidth]{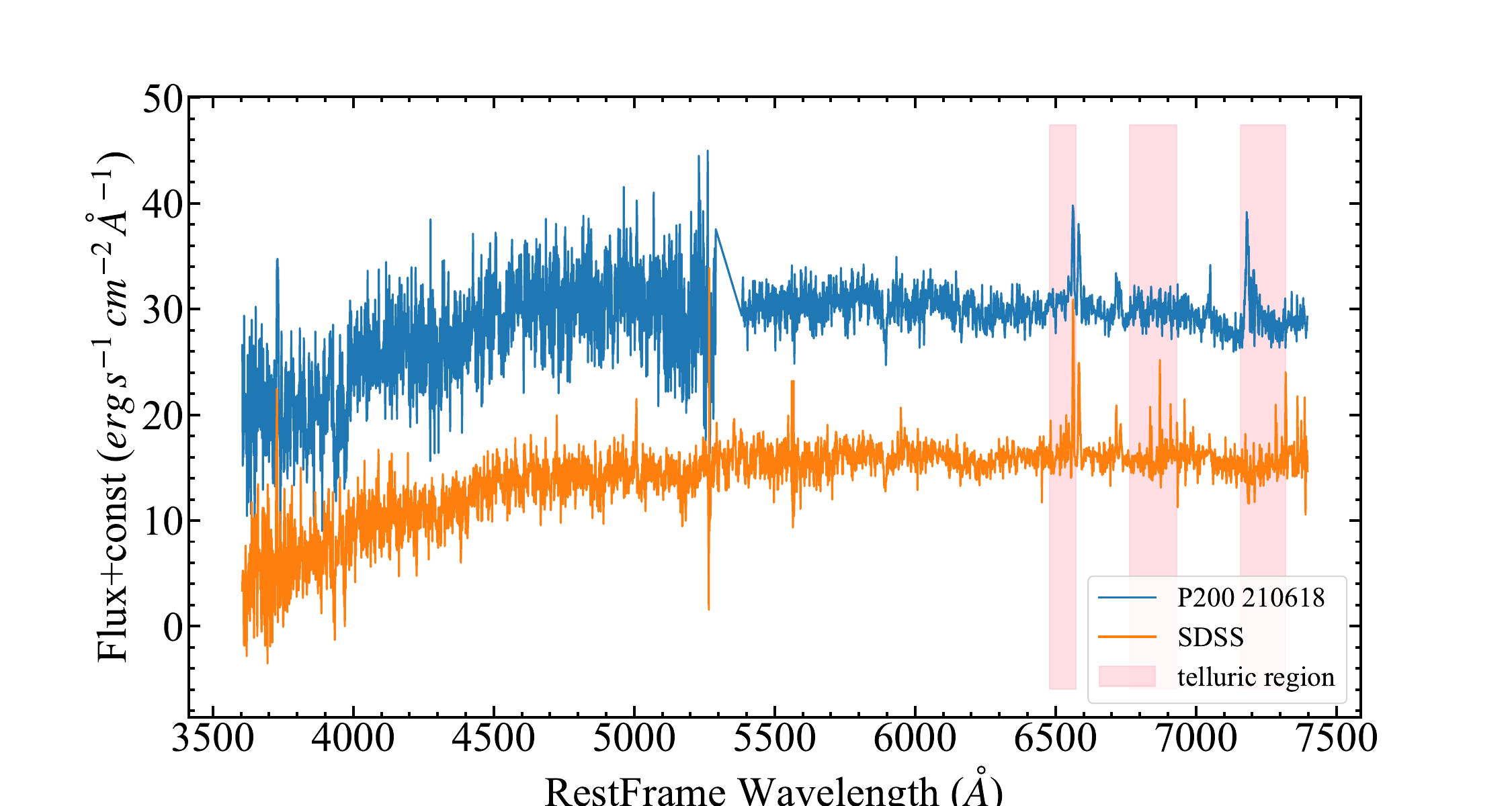}}
\end{minipage}
\caption{The optical spectra of SDSSJ1048+1228 and SDSSJ1649+2625.\label{spec}. The archival SDSS spectrum are shown in orange while the new spectra taken after the outburst is shown in blue. No obvious spectral evolution is found for both objects.
}
\end{figure}

Following the measurements performed on the entire MIRONG sample~\citep{Jiang2021apjs}, we have fitted the MIR emission of SDSSJ1048+1228 and SDSSJ1649+2625 with blackbody model and found a peak luminosity of $10^{42.94}$ and $10^{43.29}$~\lum, respectively. Such high IR luminosities almost exclude the supernova scenario and suggest that they are instead due to dust echoes of transient accretions onto SMBHs~\citep{Jiang2019,Jiang2021apjs}.
Additionally, the integrated energies radiated in IR of the two sources are $10^{50.72}$ and $10^{51.02}$~erg as of the end of 2022, which are already comparable to the most energetic supernova (e.g., SN2016aps, \citealt{Nicholl2020}), but are rather typical for transient accretions such as TDEs~\citep{vV2021c}. Furthermore, both galaxies exhibit little or weak AGN activities according to their pre-outburst SDSS spectra (Figure~\ref{spec}), and have been classified as a LINER and a composite galaxy in the Baldwin-Phillips-Terlevich (BPT) diagram~\citep{Baldwin1981}, respectively. It is worth noting that their MIR light curves show subtle variability (see Figure~\ref{lc}), This is particularly evident in the case of SDSSJ1048+1228, which suggests the possible presence of weak AGNs.
 \citet{Frederick2021} have recently reported the discovery a new class of changing-look LINERs, which pose a challenge in distinguishing them from TDEs for LINER galaxies.
We conducted spectroscopic follow-up observations with the Double spectrograph (DBSP) mounted on the Hale 200 inch telescope~\citep{Oke1982} at Palomar observatory for SDSSJ1649+2625 on 2021 June 18 and with YFOSC of the LiJiang 2.4m telescope~\citep{Fan2015-2m4,Wang2019-2m4} at Yunnan observatories for SDSSJ1048+1228 on 2023 February 18. However, no obvious spectral evolution, i.e., emerging characteristic emission lines as seen in some of the MIRONG objects~\citep{Wang2022apjs}, was found for both (see Figure~\ref{spec}). Therefore, we conclude that the two neutrino-matched MIR outbursts are highly consistent with the obscured counterparts of transient accretion onto SMBHs, such as TDEs or changing-look LINERs.

\section{Summary and Discussion}

As a rare type of nuclear transient, TDEs have been regularly discovered by optical surveys and found being a possible production site of high-energy neutrinos over past few years. Interestingly, the initial two claimed sources, one robust TDE AT2019dsg and the other TDE candidate AT2019fdr, were both accompanied by bright IR echoes in stark contrast to normal optical TDEs. This indicates that the dust in the vicinity of SMBHs, which is assumed to be responsible for the reprocessed IR emission, could be a crucial factor in producing high-energy neutrinos. The unexpected neutrino-dust link then led to the identification of a third similar candidate, AT2019aalc. Moreover, the neutrino arrivals show great consistency with IR peaks, while they are hundreds of days delayed relative to the optical peak. The intriguing correlation between high-energy neutrinos and TDEs in dusty environments has motivated us to perform a new spatial and temporal cross-match of our own MIRONG sample, which is mainly caused by transient accretion onto SMBHs, with known gold alerts of high-energy neutrino events up to the end of 2022. This cross-match has resulted in the discovery of two new candidates, SDSSJ1048+1228 and SDSSJ1649+225. Despite exhibiting huge MIR flares, these candidates show non-detection in ZTF light curves and can be reasonably categorized as obscured TDE candidates. The probability of finding two such candidates by chance is about 3\% based on our Monte Carlo simulations by shuffling the MIRONG sources in the SDSS footprint. The chance probability is much lower, that is about 0.1\% ($\sim3\sigma$), if we only consider sources with a increased flux comparable with the matched two ($W2<12.6$). Alternatively, if we restrict sources to those with a similar variability amplitude ($\Delta W2>1$), the chance probability is slightly larger, that is 0.2\%.

One may wonder whether there are more matches if taking into consideration of bronze neutrinos. We have repeated the cross-match procedure and discovered two more matches: SDSSJ104832.79+122857.2 with IceCube-200620A and SDSSJ141249.70+151254.7 with IceCube-220205A. It is worth noting that SDSSJ104832.79+122857.2 coincides with both a gold neutrino and a bronze neutrino, indicating the possibility of it being a repeating neutrino emitter as the two neutrinos are separated by a five-month interval. In addition, the new source SDSSJ141249.70+151254.7 at $z=0.1411$ shows a large variability amplitude ($\Delta W2=1.33$) yet with a lower increased flux ($W2=13.18$) due to a higher redshift in comparison with the formal two matched sources. The chance probability of matching at least four neutrinos with the 30 $\Delta W2>1$ sources keeps as low as $\sim0.1\%$. In the future, we plan to employ a more advanced likelihood-based analysis that takes into account the signalness of the neutrinos, assigning less weight to bronze events. 
 In addition, the updated 90\% containment for neutrinos up to 2020~\citep{Abbasi2023} can be also used for future refined analysis. This approach will enable us to obtain more accurate results.
 
Our findings increase the total number of TDE or TDE candidates associated with gold neutrinos to five. They all show prominent dust IR echoes and indicate strongly a possible physical connection.
The physics of producing high-energy neutrinos in TDEs remains poorly understood. Recently, \citet{Winter2022} developed time-dependent models to interpret the neutrino delays of AT2019dsg, AT2019fdr, and AT2019aalc, assuming that they are the consequence of physical scales of the postdisruption system rather than a statistical effect. They considered three models where neutrinos arise from the interactions of accelerated protons of moderate, medium, and ultra-high energies with X-rays, optical/UV, and IR photons, respectively. In the scenario of dust IR emission serving as the photons, the neutrino delay can be naturally explained by the delayed IR echo at the cost of requiring very high proton energies. In contrast, the optical/UV and X-ray photon scenario cannot describe the observed neutrino time delay as well, while they have higher neutrino production efficiency.

Our work once again highlights the importance of IR time-domain astronomy in studying nuclear transients (see also \citealt{Jiang2021apjs}), which may provide insights into the mysterious origin of high-energy neutrinos. 
It should be noted that our parent sample is constructed from flux-limited SDSS spectroscopic galaxies with a magnitude limit of $r<17.77$~\citep{Strauss2002}.  As a result, we may have missed some TDE counterparts associated with fainter photometric galaxies in our search. For example, the second event AT2019fdr, which was matched with neutrinos and has a redshift of $z=0.267$, is situated within the SDSS coverage area. However, its faintness ($r=19.17$) has prevented it from being observed spectroscopically.
To address this, we are conducting a more comprehensive blind search for TDE-like IR transients in the known neutrino sky regions using time-resolved WISE images (Z. Y. Zhou et al. in preparation). On the other hand, the persistent optical surveys conducted by ZTF~\citep{Yao2023}, along with upcoming more advanced surveys such as the Legacy Survey of Space and Time (LSST; \citealt{Ivezic2019}) and the Wide-Field Survey Telescope (WFST; \citealt{Lin2022,wfst2023}), will enable us to build a larger sample of TDEs with IR echoes. This will allow us to arrive at a more conclusive conclusion regarding whether or not dusty TDEs are a primary source of high-energy neutrinos. Lastly, it is worth emphasizing that high-energy neutrinos themselves can serve as an independent probe of obscured TDEs, once their connection has been convincingly established. \\
\\
We sincerely thank the referee for his/her very positive and constructive comments, which help improve our manuscript significantly.
We thank useful discussions with Donglian Xu on the declinational dependence of IceCube neutrino detection. This work is supported by SKA Fast Radio Burst and High-Energy Transients Project (2022SKA0130102), the National Natural Science Foundation of China (grants 11833007, 12073025, 12192221), the Fundamental Research Funds for the Central Universities (WK3440000006), and the 111 Project for "Observational and Theoretical Research on Dark Matter and Dark Energy" (B23042). The authors acknowledge the support of Cyrus Chun Ying Tang Foundations. This research uses data obtained through the Telescope Access Program (TAP). Observations obtained with the Hale Telescope at Palomar Observatory were obtained as part of an agreement between the National Astronomical Observatories, Chinese Academy of Sciences, and the California Institute of Technology. We acknowledge the support of the staff of the Lijiang 2.4m telescope. Funding for the telescope has been provided by Chinese Academy of Sciences and the People's Government of Yunnan Province”. The ZTF forced-photometry service was funded under the Heising-Simons Foundation grant \#12540303 (PI: Graham). \\

\begin{appendices} 
\onecolumngrid
\section{Appendix} 
\setcounter{table}{0}
\renewcommand{\thetable}{A\arabic{table}}
The properties of neutrinos (Table~\ref{tb-neut}) and MIRONG sources used in crossmatch (Table~\ref{tb-mirong}).

\startlongtable
\begin{deluxetable*}{lcccccccc}
\tablecaption{The Information of Neutrinos}
\label{tb-neut}
\tablehead{
\colhead{ID} & \colhead{Date} & \colhead{Time}  & \colhead{Type} & \colhead{R.A.} & \colhead{Decl.} & \colhead{ERRER90} &  \colhead{Energy} & \colhead{Signaless}
}
\colnumbers
\startdata
 1 & 22/06/24 & 16:13:16.40 & GOLD & 224.1200 &   41.3100 &  109.80 &  192.97 &  0.60870 \\
 2 & 22/05/13 & 23:23:12.60 & GOLD & 224.0300 &  -1.34000 &  62.400 &  207.92 &  0.56014 \\
 3 & 22/05/09 & 18:19:04.11 & GOLD & 334.2500 &   5.37990 &  100.79 &  176.81 &  0.44648 \\
 4 & 22/04/25 & 02:44:57.81 & GOLD & 268.2400 &  -10.7300 &  102.59 &  603.95 &  0.16877 \\
 5 & 22/04/24 & 01:06:24.06 & GOLD & 346.1100 &   8.91000 &  68.390 &  183.99 &  0.49659 \\
 6 & 22/03/06 & 03:46:37.06 & GOLD & 314.8199 &   8.60990 &  31.200 &  413.05 &  0.77357 \\
 7 & 22/03/03 & 18:00:07.62 & GOLD & 267.8000 &   11.4199 &  70.790 &  398.11 &  0.76419 \\
 8 & 22/02/05 & 20:08:10.60 & GOLD & 266.8044 &  -3.57495 &  30.800 &  215.88 &  0.59235 \\
 9 & 22/02/02 & 11:48:38.58 & GOLD & 21.35990 &  -3.87990 &  43.790 &  150.94 &  0.20564 \\
10 & 21/11/17 & 03:50:57.18 & GOLD & 225.9338 & -0.201600 &  30.800 &  195.03 &  0.52590 \\
11 & 21/09/22 & 18:17:20.94 & GOLD & 60.72990 &  -4.17990 &  39.000 &  750.76 &  0.92534 \\
12 & 21/08/11 & 02:02:44.03 & GOLD & 270.7900 &   25.2800 &  56.160 &  217.67 &  0.65826 \\
13 & 21/02/13 & 18:40:24.51 & GOLD & 155.2582 &  -35.3994 &  71.400 &  1450.4 &  0.60730 \\
14 & 21/02/10 & 11:53:55.64 & GOLD & 206.0600 &   4.78000 &  52.800 &  287.41 &  0.65464 \\
15 & 20/12/22 & 00:56:16.14 & GOLD & 206.3700 &   13.4399 &  39.600 &  185.78 &  0.53370 \\
16 & 20/12/21 & 12:36:53.45 & GOLD & 261.6899 &   41.8100 &  109.20 &  174.54 &  0.56429 \\
17 & 20/12/09 & 10:15:43.94 & GOLD & 6.860000 &  -9.25000 &  65.400 &  418.60 &  0.19226 \\
18 & 20/11/30 & 20:21:46.47 & GOLD & 30.53990 &  -12.0999 &  70.790 &  203.47 &  0.14696 \\
19 & 20/11/15 & 16:01:42.96 & GOLD & 148.7519 &  -21.6418 &  51.890 &  26182. &  0.49261 \\
20 & 20/11/15 & 02:07:26.21 & GOLD & 195.1200 &   1.37990 &  77.400 &  177.38 &  0.45973 \\
21 & 20/11/14 & 15:05:31.96 & GOLD & 105.2500 &   6.04990 &  64.800 &  214.29 &  0.56208 \\
22 & 20/10/07 & 22:01:49.28 & GOLD & 265.1700 &   5.33990 &  24.000 &  682.65 &  0.88552 \\
23 & 20/09/29 & 17:48:36.83 & GOLD & 29.51990 &   3.47000 &  31.800 &  182.89 &  0.47479 \\
24 & 20/09/26 & 07:54:11.62 & GOLD & 96.45990 &  -4.33000 &  39.600 &  670.50 &  0.44137 \\
25 & 20/07/28 & 08:17:51.99 & GOLD & 117.5554 &  -24.8475 &  30.800 &  42024. &  0.39310 \\
26 & 20/06/15 & 14:49:17.37 & GOLD & 142.9499 &   3.66000 &  73.190 &  496.36 &  0.82832 \\
27 & 20/05/30 & 07:54:29.43 & GOLD & 255.3700 &   26.6099 &  160.19 &  82.186 &  0.59170 \\
28 & 20/01/09 & 23:41:39.93 & GOLD & 164.4900 &   11.8699 &  174.60 &  375.23 &  0.76931 \\
29 & 19/10/01 & 20:09:18.17 & GOLD & 314.0799 &   12.9399 &  177.00 &  217.42 &  0.58898 \\
30 & 19/09/22 & 23:03:55.56 & GOLD & 5.759900 &  -1.57000 &  64.800 &  187.37 &  0.50501 \\
31 & 19/09/22 & 09:42:45.62 & GOLD & 167.4300 &  -22.3900 &  177.00 &  3113.9 &  0.20165 \\
32 & 19/07/30 & 20:50:41.31 & GOLD & 225.7899 &   10.4700 &  71.100 &  298.81 &  0.67158 \\
33 & 19/06/19 & 13:14:18.04 & GOLD & 343.2599 &   10.7300 &  162.59 &  198.70 &  0.54551 \\
\enddata
\tablecomments{
(2)-(3): the Date (yy/mm/dd), and the Time (hh:mm:ss.ss) of the neutrinos.
(4): the NoticeType of the neutrino (GOLD or BRONZE).
(5)-(6): R.A. and Decl. location of the neutrino (J2000 epoch) in units of degrees.
(7): the location uncertainty (radius, 90\% containment) in units of arcmin.
(8): the Eneregy most probable neutrino energy that would have produced an event with these observed parameters.
(9): the probability this is an astrophysical signal relative to backgrounds.
}
\end{deluxetable*}

\startlongtable
\begin{deluxetable*}{lccccccccc}
\tablecaption{The Information of MIRONG objects}
\label{tb-mirong}
\tablehead{
\colhead{ID} & \colhead{Name} & \colhead{R.A.}  & \colhead{Decl.} & \colhead{z} & \colhead{MJD} & \colhead{$\Delta$W1} &  \colhead{$\Delta$W2} & \colhead{W1} & \colhead{W2}
}
\colnumbers
\startdata
   1 &  SDSSJ001249.53+323233.2 &   3.2064 & 32.5426 & 0.1141 & 59566 & 0.63 & 0.93 &  13.53 &  12.64 \\
   2 &  SDSSJ002943.36-092233.8 &   7.4306 & -9.3760 & 0.0736 & 59551 & 0.21 & 0.52 &  15.69 &  14.31 \\
   3 &  SDSSJ013455.28-084352.5 &  23.7303 & -8.7312 & 0.0923 & 59200 & 0.40 & 0.55 &  14.10 &  13.03 \\
   4 &  SDSSJ013617.56+282411.3 &  24.0732 & 28.4032 & 0.0707 & 59420 & 0.31 & 0.50 &  14.76 &  13.90 \\
   5 &  SDSSJ021640.95-033747.1 &  34.1706 & -3.6298 & 0.1609 & 59578 & 0.90 & 1.20 &  14.19 &  13.29 \\
   6 &  SDSSJ030039.71-082053.8 &  45.1655 & -8.3483 & 0.0703 & 58699 & 0.27 & 0.53 &  14.52 &  13.54 \\
   7 &  SDSSJ035748.65-052239.9 &  59.4528 & -5.3778 & 0.1127 & 59081 & 0.35 & 0.57 &  14.38 &  13.32 \\
   8 &  SDSSJ075544.35+192336.4 & 118.9348 & 19.3934 & 0.1083 & 59664 & 0.33 & 0.52 &  14.68 &  13.67 \\
   9 &  SDSSJ081437.37+510141.8 & 123.6557 & 51.0283 & 0.0625 & 59504 & 0.81 & 1.49 &  13.92 &  12.75 \\
  10 &  SDSSJ082206.72+273342.6 & 125.5280 & 27.5618 & 0.0965 & 59147 & 0.35 & 0.64 &  14.40 &  13.22 \\
  11 &  SDSSJ082358.68+420259.1 & 125.9945 & 42.0498 & 0.1256 & 59665 & 0.56 & 0.85 &  15.09 &  14.21 \\
  12 &  SDSSJ082610.41+334457.2 & 126.5434 & 33.7492 & 0.1627 & 59510 & 0.93 & 1.37 &  14.29 &  13.26 \\
  13 &  SDSSJ083340.16+324759.5 & 128.4173 & 32.7999 & 0.0513 & 59669 & 0.12 & 0.54 &  15.89 &  13.75 \\
  14 &  SDSSJ083653.41+061226.0 & 129.2225 &  6.2072 & 0.1033 & 58948 & 0.41 & 0.64 &  15.15 &  14.22 \\
  15 &  SDSSJ084652.32+352741.7 & 131.7180 & 35.4616 & 0.1624 & 59671 & 0.34 & 0.51 &  14.76 &  13.81 \\
  16 &  SDSSJ090811.40+262658.7 & 137.0475 & 26.4496 & 0.2076 & 59313 & 0.39 & 0.87 &  14.58 &  13.30 \\
  17 &  SDSSJ091308.53+350003.6 & 138.2856 & 35.0010 & 0.1149 & 58948 & 0.25 & 0.62 &  15.21 &  13.81 \\
  18 &  SDSSJ091901.51+363536.1 & 139.7563 & 36.5934 & 0.1897 & 58948 & 0.30 & 0.54 &  14.95 &  13.71 \\
  19 &  SDSSJ095137.27+341612.4 & 147.9053 & 34.2701 & 0.1322 & 59687 & 0.84 & 1.08 &  14.31 &  13.49 \\
  20 &  SDSSJ100809.05+154951.4 & 152.0377 & 15.8309 & 0.1177 & 59538 & 0.56 & 0.88 &  14.03 &  12.94 \\
  21 &  SDSSJ100842.62+150409.2 & 152.1776 & 15.0692 & 0.2783 & 59696 & 0.86 & 1.14 &  14.42 &  13.43 \\
  22 &  SDSSJ101008.86+300252.6 & 152.5369 & 30.0479 & 0.0874 & 59691 & 0.31 & 0.73 &  15.70 &  14.40 \\
  23 &  SDSSJ101107.94+194428.5 & 152.7830 & 19.7413 & 0.0289 & 59173 & 0.33 & 0.79 &  13.47 &  12.13 \\
  24 &  SDSSJ101157.62+534857.8 & 152.9901 & 53.8161 & 0.2344 & 59157 & 0.46 & 0.73 &  14.72 &  13.50 \\
  25 &  SDSSJ101348.31+180734.9 & 153.4513 & 18.1264 & 0.0120 & 59696 & 0.55 & 0.50 &  14.02 &  13.94 \\
  26 &  SDSSJ101708.95+122412.2 & 154.2873 & 12.4034 & 0.1076 & 59699 & 0.60 & 0.98 &  14.84 &  13.83 \\
  27 &  SDSSJ102038.51+243708.8 & 155.1605 & 24.6191 & 0.1894 & 59331 & 0.85 & 0.95 &  13.36 &  12.58 \\
  28 &  SDSSJ102713.79+150533.2 & 156.8074 & 15.0925 & 0.1839 & 59178 & 1.24 & 1.32 &  13.75 &  13.03 \\
  29 &  SDSSJ104832.79+122857.2 & 162.1366 & 12.4825 & 0.0537 & 58977 & 0.59 & 1.11 &  13.65 &  12.60 \\
  30 &  SDSSJ110138.83+285039.2 & 165.4119 & 28.8442 & 0.0344 & 59338 & 0.37 & 0.54 &  14.27 &  13.55 \\
  31 &  SDSSJ111156.67+361707.3 & 167.9861 & 36.2854 & 0.0786 & 58970 & 0.34 & 0.68 &  14.44 &  13.18 \\
  32 &  SDSSJ111431.84+405613.8 & 168.6327 & 40.9372 & 0.1525 & 58968 & 0.53 & 0.60 &  14.02 &  13.11 \\
  33 &  SDSSJ111614.66+061736.2 & 169.0611 &  6.2934 & 0.0754 & 59194 & 0.71 & 1.04 &  13.61 &  12.80 \\
  34 &  SDSSJ112108.22+505219.6 & 170.2842 & 50.8721 & 0.1513 & 59330 & 0.51 & 0.74 &  14.82 &  13.54 \\
  35 &  SDSSJ112526.86+070520.3 & 171.3619 &  7.0890 & 0.0957 & 58828 & 0.32 & 0.59 &  15.10 &  14.03 \\
  36 &  SDSSJ112858.33+210257.5 & 172.2431 & 21.0493 & 0.1619 & 59555 & 0.42 & 0.60 &  15.08 &  14.25 \\
  37 &  SDSSJ112939.08+365301.2 & 172.4128 & 36.8837 & 0.1995 & 59182 & 0.83 & 1.10 &  15.10 &  14.18 \\
  38 &  SDSSJ115205.33+485050.0 & 178.0222 & 48.8472 & 0.1510 & 59179 & 0.96 & 1.21 &  13.66 &  12.73 \\
  39 &  SDSSJ115812.34+384332.3 & 179.5514 & 38.7257 & 0.0742 & 59344 & 0.25 & 0.75 &  15.72 &  14.07 \\
  40 &  SDSSJ120344.13+543331.7 & 180.9339 & 54.5588 & 0.1052 & 59334 & 0.51 & 0.85 &  14.86 &  13.94 \\
  41 &  SDSSJ120444.71+104642.3 & 181.1863 & 10.7784 & 0.0685 & 59566 & 0.75 & 1.19 &  12.55 &  11.26 \\
  42 &  SDSSJ120813.78+513504.2 & 182.0574 & 51.5845 & 0.0889 & 59179 & 0.21 & 0.75 &  15.89 &  14.06 \\
  43 &  SDSSJ121249.92+024012.9 & 183.2080 &  2.6703 & 0.0772 & 59571 & 0.22 & 0.63 &  16.33 &  14.89 \\
  44 &  SDSSJ121337.51+051730.8 & 183.4063 &  5.2919 & 0.1398 & 58842 & 0.24 & 0.52 &  15.79 &  14.66 \\
  45 &  SDSSJ121738.27+035040.5 & 184.4094 &  3.8447 & 0.0731 & 59207 & 0.37 & 0.75 &  14.37 &  13.31 \\
  46 &  SDSSJ121825.52+295154.8 & 184.6063 & 29.8652 & 0.1356 & 59197 & 0.89 & 1.57 &  14.23 &  12.82 \\
  47 &  SDSSJ123155.15+323240.4 & 187.9798 & 32.5445 & 0.0654 & 59721 & 1.14 & 1.44 &  12.20 &  11.42 \\
  48 &  SDSSJ123431.24+235338.4 & 188.6302 & 23.8942 & 0.2787 & 58995 & 0.45 & 0.61 &  14.90 &  14.06 \\
  49 &  SDSSJ124754.96-033738.7 & 191.9790 & -3.6274 & 0.0903 & 59216 & 1.14 & 1.48 &  11.78 &  10.87 \\
  50 &  SDSSJ125953.97+060050.3 & 194.9749 &  6.0140 & 0.1060 & 59011 & 0.41 & 0.65 &  14.61 &  13.61 \\
  51 &  SDSSJ130617.74+533907.1 & 196.5739 & 53.6520 & 0.0300 & 59555 & 0.40 & 0.74 &  13.23 &  12.23 \\
  52 &  SDSSJ132338.83+275420.2 & 200.9118 & 27.9056 & 0.0730 & 58847 & 0.55 & 1.43 &  13.88 &  12.12 \\
  53 &  SDSSJ134148.78+370047.2 & 205.4533 & 37.0131 & 0.1968 & 59006 & 0.50 & 0.73 &  14.68 &  13.68 \\
  54 &  SDSSJ134419.60+512624.8 & 206.0816 & 51.4402 & 0.0629 & 59199 & 0.58 & 0.77 &  12.69 &  11.90 \\
  55 &  SDSSJ134821.40+024914.9 & 207.0892 &  2.8208 & 0.0855 & 59593 & 0.77 & 1.29 &  13.53 &  12.49 \\
  56 &  SDSSJ134958.72-001120.9 & 207.4947 & -0.1891 & 0.1020 & 59595 & 0.32 & 0.55 &  15.18 &  14.20 \\
  57 &  SDSSJ135246.87+202444.0 & 208.1953 & 20.4122 & 0.0544 & 59223 & 0.66 & 1.19 &  13.88 &  12.86 \\
  58 &  SDSSJ135251.12+161317.2 & 208.2130 & 16.2214 & 0.1578 & 58858 & 0.51 & 0.82 &  14.79 &  13.76 \\
  59 &  SDSSJ140515.60+542458.1 & 211.3150 & 54.4161 & 0.0833 & 59359 & 0.45 & 0.71 &  14.84 &  13.99 \\
  60 &  SDSSJ141249.70+151254.7 & 213.2071 & 15.2152 & 0.1411 & 59594 & 1.01 & 1.33 &  13.96 &  13.18 \\
  61 &  SDSSJ141437.38-002800.8 & 213.6557 & -0.4669 & 0.2611 & 59236 & 0.18 & 0.56 &  16.67 &  14.70 \\
  62 &  SDSSJ142237.73+284424.3 & 215.6572 & 28.7401 & 0.1250 & 59384 & 0.64 & 1.17 &  14.26 &  13.01 \\
  63 &  SDSSJ142420.79+624916.4 & 216.0866 & 62.8212 & 0.1091 & 59713 & 0.39 & 0.62 &  14.32 &  13.40 \\
  64 &  SDSSJ142727.72+162347.5 & 216.8655 & 16.3965 & 0.1278 & 59233 & 1.05 & 1.50 &  14.01 &  13.26 \\
  65 &  SDSSJ142813.88+391218.6 & 217.0578 & 39.2052 & 0.2583 & 58651 & 0.60 & 0.62 &  14.08 &  13.37 \\
  66 &  SDSSJ143016.05+230344.4 & 217.5668 & 23.0623 & 0.0810 & 58866 & 0.68 & 0.93 &  12.62 &  11.78 \\
  67 &  SDSSJ143114.72+171135.4 & 217.8113 & 17.1932 & 0.0378 & 58869 & 0.81 & 1.22 &  15.23 &  14.24 \\
  68 &  SDSSJ143701.26+073508.4 & 219.2553 &  7.5857 & 0.1830 & 59238 & 0.84 & 1.15 &  13.69 &  12.53 \\
  69 &  SDSSJ143701.72+532358.4 & 219.2572 & 53.3996 & 0.0469 & 59208 & 0.42 & 0.75 &  13.85 &  13.01 \\
  70 &  SDSSJ144227.60+555846.4 & 220.6150 & 55.9796 & 0.0769 & 59205 & 1.36 & 1.83 &  11.67 &  10.81 \\
  71 &  SDSSJ145501.50+084427.8 & 223.7562 &  8.7411 & 0.1601 & 58878 & 0.29 & 0.73 &  15.62 &  13.94 \\
  72 &  SDSSJ150440.39+010735.0 & 226.1683 &  1.1264 & 0.1283 & 59613 & 0.67 & 0.75 &  12.37 &  11.35 \\
  73 &  SDSSJ151828.07+191636.6 & 229.6170 & 19.2768 & 0.0894 & 59244 & 0.28 & 0.55 &  15.43 &  14.42 \\
  74 &  SDSSJ152223.27+014626.1 & 230.5970 &  1.7739 & 0.0785 & 59253 & 0.41 & 0.61 &  14.38 &  13.67 \\
  75 &  SDSSJ152535.07+082157.5 & 231.3961 &  8.3660 & 0.0754 & 59616 & 0.42 & 0.99 &  15.73 &  14.35 \\
  76 &  SDSSJ153239.62+443203.1 & 233.1651 & 44.5342 & 0.0376 & 59234 & 1.00 & 1.76 &  13.54 &  12.40 \\
  77 &  SDSSJ154357.60+100613.5 & 235.9900 & 10.1037 & 0.1753 & 59256 & 0.92 & 1.42 &  14.37 &  13.33 \\
  78 &  SDSSJ154843.07+220812.7 & 237.1795 & 22.1369 & 0.0313 & 59253 & 2.56 & 3.48 &  11.18 &  10.16 \\
  79 &  SDSSJ154849.49+125513.9 & 237.2062 & 12.9205 & 0.1433 & 59418 & 0.53 & 0.93 &  14.97 &  13.71 \\
  80 &  SDSSJ155259.95+210246.9 & 238.2498 & 21.0463 & 0.1717 & 59417 & 0.96 & 1.24 &  14.69 &  13.75 \\
  81 &  SDSSJ155440.26+362952.0 & 238.6678 & 36.4978 & 0.2368 & 59246 & 0.88 & 1.21 &  14.23 &  13.26 \\
  82 &  SDSSJ155633.99+174200.8 & 239.1416 & 17.7002 & 0.1359 & 59419 & 0.43 & 0.62 &  15.22 &  14.31 \\
  83 &  SDSSJ160014.59+180400.0 & 240.0608 & 18.0667 & 0.1381 & 59053 & 0.27 & 0.51 &  15.49 &  14.35 \\
  84 &  SDSSJ161258.17+141617.6 & 243.2424 & 14.2715 & 0.0720 & 59262 & 0.43 & 0.88 &  14.75 &  13.34 \\
  85 &  SDSSJ164938.77+262515.3 & 252.4116 & 26.4211 & 0.0588 & 59068 & 1.41 & 2.29 &  13.09 &  11.92 \\
  86 &  SDSSJ165516.63+382802.5 & 253.8193 & 38.4674 & 0.1005 & 58699 & 0.39 & 0.65 &  15.36 &  14.43 \\
  87 &  SDSSJ170503.58+344010.3 & 256.2649 & 34.6695 & 0.1659 & 58706 & 0.18 & 0.58 &  16.55 &  14.89 \\
  88 &  SDSSJ214906.34+122519.6 & 327.2764 & 12.4221 & 0.0596 & 58791 & 0.28 & 0.59 &  14.98 &  13.53 \\
  89 &  SDSSJ215055.73-010654.2 & 327.7322 & -1.1151 & 0.0879 & 58990 & 1.79 & 1.96 &  13.83 &  12.94 \\
  90 &  SDSSJ221541.60-010721.1 & 333.9233 & -1.1225 & 0.0478 & 58996 & 0.24 & 0.68 &  14.17 &  12.75 \\
  91 &  SDSSJ223628.13+142830.9 & 339.1172 & 14.4753 & 0.2649 & 59738 & 0.25 & 0.58 &  15.97 &  14.63 \\
  92 &  SDSSJ225549.89+001728.5 & 343.9579 &  0.2913 & 0.1111 & 59372 & 0.29 & 0.89 &  15.87 &  14.14 \\
  93 &  SDSSJ233517.46+010421.1 & 353.8227 &  1.0725 & 0.0806 & 59745 & 0.42 & 0.78 &  14.91 &  13.89 \\
  94 &  SDSSJ235911.19+011006.8 & 359.7966 &  1.1686 & 0.0985 & 58655 & 0.79 & 0.96 &  13.71 &  12.98 \\
\enddata
\tablecomments{
(1): the ID of MIRONG sources with W2 peak time in the same range of neutrinos listed in Appendix Table~\ref{tb-neut}, that is between June 2019 June and June 2022.
(2)-(5): the Name, R.A., Decl. and redshift of these objects.
(6): the Modified Julian Date (MJD) of the W2 peak. 
(7)-(8): the variability amplitudes in W1 and W2 band, respectively.
(9)-(10): the W1 and W2 magnitudes with flux of quiescent state subtracted.
}
\end{deluxetable*}

\end{appendices}

\end{document}